\documentclass[aps,pre,superscriptaddress,onecolumn,amsmath,amssymb,showpacs]{revtex4}
\usepackage{graphicx}
\usepackage{dcolumn}
\usepackage{bm}
\usepackage[bookmarks=false,dvips]{hyperref}
\hypersetup{colorlinks,citecolor=blue, linkcolor=blue, urlcolor=blue, anchorcolor=blue}

\begin{document}


\title{Mean Trajectories of Multiple Tracking Points on A Brownian Rigid Body: Convergence, Alignment and Twist}


\author{Jianping Xu}
  \email{xujp@utexas.edu}
  \affiliation{The University of Texas at Austin, Austin, Texas 78712, USA}

\date{\today}

\begin{abstract}
We consider mean trajectories of multiple tracking points on a rigid body that conducts Brownian motion in the absence and presence of an external force field. Based on a na\"{\i}ve representation of rigid body - polygon and polyhedron where hydrodynamic interactions are neglected, we study the Langevin dynamics of these Brownian polygons and polyhedra. Constant force, harmonic force and an exponentially decaying force are investigated as examples. In two dimensional space, depending on the magnitude and form of the external force and the isotropy and anisotropy of the body, mean trajectories of these tracking points can exhibit three regimes of interactions: convergence, where the mean trajectories converge to either a point or a single trajectory; alignment, where the mean trajectories juxtapose in parallel; twist, where the mean trajectories twist and intertwine, forming a plait structure. Moreover, we have shown that in general a rigid body can sample from these regimes and transit between them. And its Brownian behavior could be modified during such transition. Notably, from a polygon in two dimensional space to a polyhedron in three dimensional space, the alignment and twist regimes disappear and there is only the convergence regime survived, due to the two more rotational degrees of freedom in three dimensional space.
\end{abstract}



\maketitle


\section{Introduction}
A rigid body~\cite{favro,miguelx,delong} that conducts Brownian motion can translate and rotate in space. Most interestingly, in scenarios where the particle is screwlike~\cite{brenner1,brenner2}, L-shaped~\cite{Kummel}, biaxial~\cite{Wittkowski} and ellipsoidal~\cite{Han0}, etc., translation and rotation can couple~\cite{brenner1,brenner2,Sun,Ayan,Ayan2,Ayan3,Han0}, leading to a rich class of trajectory patterns, e.g., helical motion~\cite{Wittkowski}, circular motion~\cite{Kummel}. In recent years, apart from exploring these novel dynamic behaviors arising from rigid-body Brownian motion, there were general models built, such as Brownian Dynamics~\cite{ermak,miguelx}, Stokesian Dynamics~\cite{brady,swan}, Fluctuating Hydrodynamics~\cite{Sharma} and the Langevin dynamics of arbitrarily shaped particle~\cite{Sun,delong}. Usually in these models the effects of non-stochastic factors such as hydrodynamic interactions and particle geometry enter into the displacement equations as resistance tensor~\cite{Sun} or equivalently the mobility tensor~\cite{delong}, while stochasticity is contained in force and torque terms. Subsequently, the trajectory of a tracking point (TP) on the body, e.g., the center of mass (CoM), or the center of friction (CoF, also known as center of hydrodynamic stress~\cite{Ayan}) is generated. Hence the particle is still represented by a zero volume TP rather than a finite volume body.

However, to a large extent, the analysis of a single trajectory of a single TP could indeed provide rich information regarding the particle's physical properties and its interactions with the environment. Methods like single trajectory analysis~\cite{tejedor,holcman} and power spectral analysis~\cite{Schnellb,Sposini} can be utilized to extract useful information of the particle, e.g., diffusion coefficient~\cite{Xavier1} and mean squared displacement~\cite{Xavier2}. Most recently there is exciting new model based on information theory to infer the external force field from a stochastic trajectory~\cite{Frishman}. Indeed, the toolkit one can use to decipher a trajectory is updating fast. Nevertheless, when particle size matters the trajectory traced by a single TP does not reveal the orientation of the body along its path, thus not enough to characterize the particle's state. And even worse, trajectory recorded by tracking an inappropriately chosen TP could contain error.

\begin{figure}
\includegraphics[width=0.45\textwidth]{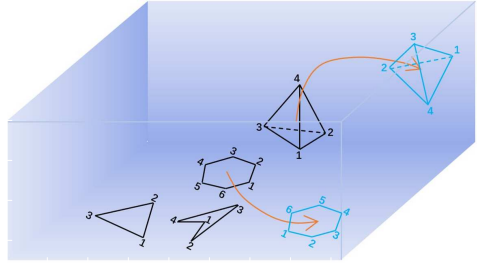}
\caption{\label{fig:pgschematics} (Color Online) Schematics of the polygon/polyhedron representation. Each labeled vertex traces a trajectory in space as the body moves through.}
\end{figure}

It is therefore natural to consider trajectories of multiple TPs on a rigid body because this helps us understand either the difference or commonality among different TPs. Since the object is Brownian, it is necessary to consider the mean trajectory. Note that to obtain the ``mean trajectory'' of a designated TP, one must consider an ensemble of identical body and average based on the ensemble. We expect to find simple but characteristic regimes of interaction among the mean trajectories of different tracking points. To achieve this, we adopt an extremely simplified representation - polygon (in 2D) and polyhedron (in 3D). As shown in Fig. \ref{fig:pgschematics}, each vertex of the polygon/polyhedron is a mass point with mass $m_i$ and friction coefficient $\xi_i$ ($1\leq i\leq n$, $n$ being the number of vertices.) Vertex $i$ experiences thermal fluctuation force $\delta\mathbf{F}_i$, friction force $\mathbf{f}_i$ exerted by the fluids and external force $\mathbf{F}_i$. Edges are assumed rigid,  massless and noninteractive with the environment. Hydrodynamic interaction among vertices (which are mediated by fluids) is also neglected. Consequently, it leads to a simple Langevin equation system which describes translation and rotation in space. This setup resembles a dot multisubunit representation~\cite{Torre4} and the bead representation~\cite{delong,Sun}, although in many previous studies the hydrodynamic interactions among beads are preserved to make more realistic cases. However, our simplified model turns out not too simplified and is able to generate rich dynamic behaviors. Besides, as shown in the Appendix, the mean displacement curve generated from this model agrees well with experimental and theoretical results of boomerang colloidal particle~\cite{Ayan}.

The paper is organized as follows. In Section \ref{model}, the construction of the Langevin dynamics model of the Brownian polygon/polyhedron system is presented. Details of computation are presented. In Section \ref{Con}, the convergence regime for motion in 2D space is identified and modeled. In Section \ref{Align}, the alignment regime for motion in 2D space is identified and modeled. In Section \ref{twist}, the twist regime for motion in 2D space is identified and modeled. In Section \ref{transition}, we discuss the transition between regimes and the modification of Brownian behavior in the transition. In Section \ref{2to3}, we extend the 2D investigations to 3D. Finally, concluding remarks are presented in Section \ref{Conclusion}.

\section{\label{model}Model Construction and Computation}
Denote the position vector of the \textit{i}-th vertex as $\mathbf{r}_i$. It is customary to define various ``centers''~\cite{Ayan} of the body. In our context, first, the geometric center (GC), whose position vector is $\mathbf{r}_c=\frac{1}{n}\sum_{i=1}^{n}\mathbf{r}_i$. Second, CoM, whose position vector is $\mathbf{r}_m=\frac{1}{\sum_{i=1}^{n}m_i}\sum_{i=1}^{n}m_i\mathbf{r}_i$. Third, CoF, whose position vector is $\mathbf{r}_f=\frac{1}{\sum_{i=1}^{n}\xi_i}\sum_{i=1}^{n}\xi_i\mathbf{r}_i$. Depending on the distribution of mass and friction, these three centers can separate or coincide. Then, denote the vectors joining from these centers to vertex \textit{i} as $\mathbf{R}^c_i (=\mathbf{r}_i-\mathbf{r}_c)$, $\mathbf{R}^m_i (=\mathbf{r}_i-\mathbf{r}_m)$ and $\mathbf{R}^f_i (=\mathbf{r}_i-\mathbf{r}_f)$, respectively. These vectors have the properties
\begin{equation}
\sum_{i=1}^{n}\mathbf{R}^c_i=\mathbf{0},\quad\sum_{i=1}^{n}m_i\mathbf{R}^m_i=\mathbf{0},\quad\sum_{i=1}^{n}\xi_i\mathbf{R}^f_i=\mathbf{0},
\label{eq:prop1}
\end{equation}
which can be easily shown to be true. They attach to the body and can translate and rotate in the lab frame.
Motion of the polygon/polyhedron is decomposed into translation of CoM and rotation relative to CoM. Denote vertex \textit{i}'s velocity in the lab frame as $\mathbf{v}_i$. Obviously, $\mathbf{v}_i=\mathbf{v}_m+\boldsymbol{\omega}\times\mathbf{R}^m_i$, where $\mathbf{v}_m$ is the velocity of CoM, $\boldsymbol{\omega}$ the angular velocity. Newtonian mechanics of $\{\boldsymbol{\omega},\mathbf{v}_m\}$~\cite{Sun} in the polygon/polyhedron picture writes,
\begin{equation}
\begin{aligned}
\big(\sum_{i=1}^{n}m_i|\mathbf{R}^m_i|^2\big)\frac{d\boldsymbol{\omega}}{dt}&=\sum_{i=1}^{n}\mathbf{R}^m_i\times(\mathbf{f}_i+\delta\mathbf{F}_i+\mathbf{F}_i),\\
\big(\sum_{i=1}^n m_i\big)\frac{d\mathbf{v}_m}{dt}&=\sum_{i=1}^{n}(\mathbf{f}_i+\delta\mathbf{F}_i+\mathbf{F}_i),\\
\end{aligned}
\label{eq:torgue_momentum}
\end{equation}
where $\sum_{i=1}^{n}m_i|\mathbf{R}^m_i|^2$ is the moment of inertia. $|\mathbf{R}^m_i|$ is the length of the vector and stays constant. $\delta\mathbf{F}_i$ observes
\begin{equation}
\langle\delta\mathbf{F}_i(t)\delta\mathbf{F}_j(t')\rangle=2\xi_ikT\mathbf{B}\delta_{ij}\delta(t-t'),
\label{eq:fluc-dissi}
\end{equation}
where $\mathbf{B}$ is an identity matrix, $k$ is the Boltzmann constant, $T$ is the temperature, $\delta_{ij}$ is the Kronecker sign, $\delta(\cdot)$ is the Dirac delta function and $\langle\cdot\rangle$ is the ensemble average. In general, Eq. (\ref{eq:fluc-dissi}) could be written as $\langle\delta\mathbf{F}_i(t)\delta\mathbf{F}_j(t')\rangle=2kT\boldsymbol{\Xi}\delta_{ij}\delta(t-t')$~\cite{Sun}, where $\boldsymbol{\Xi}$ is a resistance tensor. In our simple representation, $\boldsymbol{\Xi}$ reduces to $\xi_i\mathbf{B}$. $\delta_{ij}$ assumes thermal fluctuation at one vertex is not correlated to that at another vertex, which is reasonable. $\mathbf{f}_i=-\xi_i\mathbf{v}_i=-\xi_i(\mathbf{v}_m+\boldsymbol{\omega}\times\mathbf{R}^m_i)$. Substituting $\mathbf{f}_i$ into Eq. (\ref{eq:torgue_momentum}), after simple algebraic manipulations one arrives at the Langevin equations,
\begin{equation}
\begin{aligned}
\big(\sum_{i=1}^{n}m_i|\mathbf{R}^m_i|^2\big)\frac{d\boldsymbol{\omega}}{dt}=&-\big(\sum_{i=1}^{n}\xi_i|\mathbf{R}^m_i|^2\big)\boldsymbol{\omega}-\sum_{i=1}^{n}\mathbf{R}^m_i\times\xi_i\mathbf{v}_m
+\sum_{i=1}^{n}\mathbf{R}^m_i\times\delta\mathbf{F}_i+\sum_{i=1}^{n}\mathbf{R}^m_i\times\mathbf{F}_i,\\
\big(\sum_{i=1}^{n}m_i\big)\frac{d\mathbf{v}_m}{dt}=&-\big(\sum_{i=1}^{n}\xi_i\big)\mathbf{v}_m-\sum_{i=1}^{n}\xi_i\boldsymbol{\omega}\times\mathbf{R}^m_i+\sum_{i=1}^{n}\delta\mathbf{F}_i+\sum_{i=1}^{n}\mathbf{F}_i.\\
\end{aligned}
\label{eq:gen-eqn}
\end{equation}

An additional equation for $\mathbf{R}^m_i$ comes from the rigidity of the body. One could write $\mathbf{R}^m_i-\mathbf{R}^m_i(0)=\Delta\mathbf{r}_i-\Delta\mathbf{r}_m=\int_{0}^{t}\mathbf{v}_idt'-\int_{0}^{t}\mathbf{v}_mdt'=\int_{0}^{t}\boldsymbol{\omega}\times\mathbf{R}^m_idt'$, where $\Delta\mathbf{r}$ is the displacement in lab frame. Equivalently,
\begin{equation}
\frac{d\mathbf{R}^m_i}{dt}=\boldsymbol{\omega}\times\mathbf{R}^m_i.
\label{eq:prop2}
\end{equation}

Eqs.~(\ref{eq:gen-eqn}-\ref{eq:prop2}) describe the Langevin dynamics for a polygon/polyhedron. Several comments about Eqs.~(\ref{eq:gen-eqn}-\ref{eq:prop2}):

a): Different polygon/polyhedron geometries result in different sets of vectors $\{\mathbf{R}^m_i\}_{1\leq i\leq n}$. On the one hand, these vectors enter the moment of inertia/friction and impact the relaxation rate of angular velocity. On the other hand, they participate in the torques acting on the body, as well as the translation-rotation coupling term. Third, geometry impacts the relative position of GC, CoM and CoF, which is important in the mean trajectories patterns. In the numerical investigations we will study different geometries.

b): Stochasticity is introduced into the system through $\delta\mathbf{F}_i$. $\delta\mathbf{F}_i=\sqrt{2\xi_ikT}\mathbf{W}(t)$, where $\mathbf{W}(t)$ is a vector and each of its component is a Gaussian white noise. Stochasticity participates in driving the evolution of velocity and angular velocity. It competes with deterministic force to shape the evolution of a Brownian trajectory~\cite{Frishman}.

c): Eq. (\ref{eq:gen-eqn}) explicitly contains translation-rotation coupling~\cite{Sun,Ayan,Ayan2,Ayan3,Han0} terms. Under such coupling, how the body translates influences how it rotates, and vice versa. However, if $\xi_i/m_i=\xi_j/m_j|_{1\leqslant i\neq j\leqslant n}$, clearly $\sum_{i=1}^{n}\xi_i\mathbf{R}^m_i=const.\cdot\sum_{i=1}^{n}m_i\mathbf{R}^m_i=\mathbf{0}$ (Eq. (\ref{eq:prop1})), hence the coupling terms vanish.  Under such condition, translation and rotation decouple. This forms a criterion of categorizing different polygons/polyhedra as translation-rotation coupled (TRC) or non-TRC, as shown in Tab. \ref{tab:isoaniso_trc}. Since $\{\xi_i\}_{1\leqslant i\leqslant n}$ measures the interaction strength between vertices and the carrying medium, we can also categorize polygons/polyhedra based on how $\xi_i$'s are distributed. We define here a body is isotropic if $\xi_i=\xi_j|_{1\leqslant i\neq j\leqslant n}$ is true, and anisotropic if it's not true. These two criteria overlap and give finer description of the body. Noteworthily, it is readily verifiable that for Eq. (\ref{eq:gen-eqn}) all non-TRC bodies should behave similarly, regardless of isotropy and anisotropy, which only make a difference in the relaxation rate. Therefore, for simplicity it suffices to investigate isotropic non-TRC, isotropic TRC and anisotropic TRC under different geometries.
\begin{table}
\caption{Categorization of a polygon/polyhedron. $A=\{\xi_i/m_i=\xi_j/m_j|_{1\leqslant i\neq j\leqslant n}\mathrm{\,is\,true}\}$, $B=\{\xi_i=\xi_j|_{1\leqslant i\neq j\leqslant n}\mathrm{\,is\,true}\}$.}
\label{tab:isoaniso_trc}
\begin{center}
\begin{tabular}{ccc}
\hline
\hline
 & $A$ & $^{\neg}A$\\
\hline
$B$ & Isotropic non-TRC & Isotropic TRC\\
$^{\neg}B$ & Anisotropic non-TRC & Anisotropic TRC\\
\hline
\hline
\end{tabular}
\end{center}
\end{table}

d): Given a geometry, $\boldsymbol{\omega}(t)$, $\mathbf{v}_m(t)$, and $\mathbf{R}^m_i(t)$ are solved numerically. Initially, we set $\boldsymbol{\omega}(0)=\mathbf{0}$ and $\mathbf{v}_m(0)=\mathbf{0}$. $\mathbf{R}^m_i(0)$ depends on the initial placement of the body in the coordinates. Then these vectors are evolved according to Eqs.~(\ref{eq:gen-eqn}-\ref{eq:prop2}). Time $t$ is discretized into $t=N\Delta t$, $N$ is the $N$-th time step and $\Delta t$ is the time step size. The simulation is carried out in a unitless fashion, $kT=1$, time step $\Delta t=0.1$. At each time step, we sample $\mathbf{W}(N)$ in $x$, $y$ (for 2D) and $x$, $y$, $z$ (for 3D) directions independently from a Gaussian probability density function of zero mean and standard deviation of $\Delta t$, such that we get $\delta\mathbf{F}_i(N)$. A simplified representation for the numerical iterations is as follows,
\begin{equation}
\begin{aligned}
&\boldsymbol{\omega}(N)=\boldsymbol{\omega}(N-1)+\Delta t\cdot f\big(\boldsymbol{\omega}(N-1),\mathbf{R}^m_i(N-1),\mathbf{v}_m(N-1),\delta\mathbf{F}_i(N-1),\mathbf{F}_i(N-1)\big);\\
&\mathbf{v}_m(N)=\mathbf{v}_m(N-1)+\Delta t\cdot g\big(\boldsymbol{\omega}(N-1),\mathbf{R}^m_i(N-1),\mathbf{v}_m(N-1),\delta\mathbf{F}_i(N-1),\mathbf{F}_i(N-1)\big);\\
&\mathbf{R}^m_i(N)=\mathbf{R}^m_i(N-1)+\Delta t\cdot h\big(\boldsymbol{\omega}(N-1),\mathbf{R}^m_i(N-1)\big);
\end{aligned}
\label{eq:numerical_scheme}
\end{equation}
where $f()$, $g()$, $h()$ are algebraic operations given by Eqs.~(\ref{eq:gen-eqn}-\ref{eq:prop2}), $N=1,2,3,4\cdots$. This is an explicit scheme after applying Euler's method and it is easy to execute. Such scheme works well if $\Delta t$ is small, e.g., in our case $\Delta t=0.1$. One may also try method such as Runge-Kutta, which is computationally more expensive but is more tolerant to a large time step. Given $\boldsymbol{\omega}(0)$, $\mathbf{v}_m(0)$ and $\mathbf{R}^m_i(0)$, Eq. (\ref{eq:numerical_scheme}) evolves the time series of these quantities. After obtaining these quantities, one could integrate to get displacement and hence the trajectory. To obtain the mean trajectory, one must run the scheme multiple times to get the ensemble average. In our computations, each mean trajectory is averaged based on 2400 realizations.

\section{\label{Con}Convergence of mean trajectories}
\begin{figure}
\includegraphics[width=0.4\textwidth]{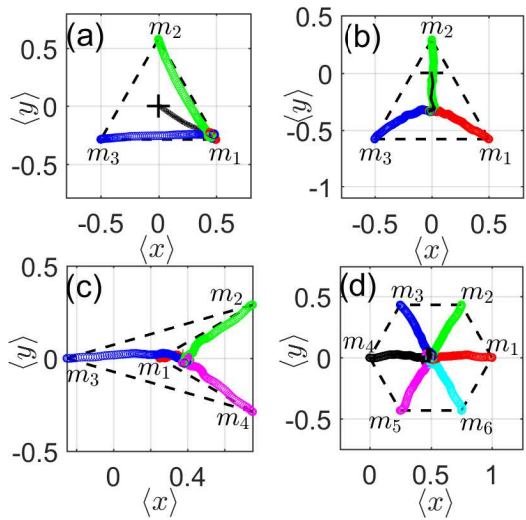}
\caption{\label{fig:mdCompare}(Color Online) Mean trajectories of vertices and CoM in the absence of external force. Black dashed lines are initial placement of the polygon. (a) Equilateral triangle, anisotropic TRC. (b) Equilateral triangle, isotropic TRC. (c) Arrow-shaped polygon and (d) equilateral hexagon, isotropic non-TRC.}
\end{figure}

The simplest situation is a freely roaming polygon in 2D space where no external force is present ($\{\mathbf{F}_i\}_{1\leq i\leq n}=\{\mathbf{0}\}$). The polygon is driven only by $\{\delta\mathbf{F}_i\}_{1\leq i\leq n}$.
Eqs. (\ref{eq:gen-eqn}-\ref{eq:prop2}) are solved under some representative geometries.
First we consider an equilateral triangle of anisotropic TRC type, as shown in Fig. \ref{fig:mdCompare} (a).
The mean trajectories of the three vertices and CoM till $t=500$ in the $\langle x\rangle$-$\langle y\rangle$ space are generated. In this case, $m_1=m_2=m_3=1$, $\xi_1=0.28$, $\xi_2=\xi_3=0.01$. CoM coincides with GC at $(0,0)$ (black cross in the figure). The CoF by definition is at $(\frac{0.9}{2},-\frac{0.9\sqrt{3}}{6})$ , which is very close to vertex 1's initial position $(\frac{1}{2},-\frac{\sqrt{3}}{6})$. Results show that the mean trajectories of the three vertices (red, green, blue circles for vertices 1, 2, 3) and the CoM (black solid line) converge to the CoF unanimously.

In Fig. \ref{fig:mdCompare} (b), isotropic TRC is considered - $m_1=m_3=0.5$, $m_2=2$, and $\xi_1=\xi_2=\xi_3=0.1$. Again, CoM is initially placed at $(0,0)$. CoF coincides GC at $(0,-\frac{\sqrt{3}}{6})$. The mean trajectories converge to CoF as well.
Fig. \ref{fig:mdCompare} (c) and (d) compare two isotropic non-TRC cases with rather different shapes - (c) an arrow-shaped polygon and in (d) an equilateral hexagon. In (c), vertices 1, 2, 3, 4's initial positions are $(\frac{1}{4},0)$, $(\frac{3}{4},\frac{\sqrt{3}}{6})$, $(-\frac{1}{4},0)$, $(\frac{3}{4},-\frac{\sqrt{3}}{6})$, respectively. $m_1=m_2=m_3=m_4=0.75$, $\xi_1=\xi_2=\xi_3=\xi_4=0.075$. CoF, GC and CoM coincide at $(\frac{3}{8},0)$. However, because of concave geometry, these centers fall out of the body~\cite{Ayan}. Again, all the trajectories converge to CoF. In (d), vertices 1, 2, 3, 4, 5, 6's initial positions are $(1,0)$, $(\frac{3}{4},\frac{\sqrt{3}}{4})$, $(\frac{1}{4},\frac{\sqrt{3}}{4})$, $(0,0)$, $(\frac{1}{4},-\frac{\sqrt{3}}{4})$, $(\frac{3}{4},-\frac{\sqrt{3}}{4})$. $m_1=m_2=m_3=m_4=m_5=m_6=0.5$, $\xi_1=\xi_2=\xi_3=\xi_4=\xi_5=\xi_6=0.05$. CoF, GC and CoM coincide at $(\frac{1}{2},0)$. In this case, the mean trajectories converge to CoF as well. This convergence behavior in the absence of external force is also verified through a semi-analytical solution to Eqs. (\ref{eq:gen-eqn}-\ref{eq:prop2}), as shown in the Appendix.

Surprisingly, the above results suggest that the convergence behavior is invariant with respect to change in polygon geometries.
However, change in geometries may lead to observational effects. A concave body's CoF can be outside the body (like in Fig. \ref{fig:mdCompare} (c)), whereas CoF of a convex body is inside. Therefore the observed Brownian motion of a convex body, such as a sphere, looks unbiased, while for a concave body, such as a boomerang particle, looks biased~\cite{Ayan}.

Given a geometry, we further show that details of convergence depend on size of the system. Based on case of Fig. \ref{fig:mdCompare} (a), we explored how the size of the system influences the spatial and temporal scale of the convergence. A comparison of MD of the CoM when $l=1$, $l=2$, $l=4$ and $l=8$ is made, where $l$ is the edge length of the triangle.
\begin{figure}
\includegraphics[width=0.7\textwidth]{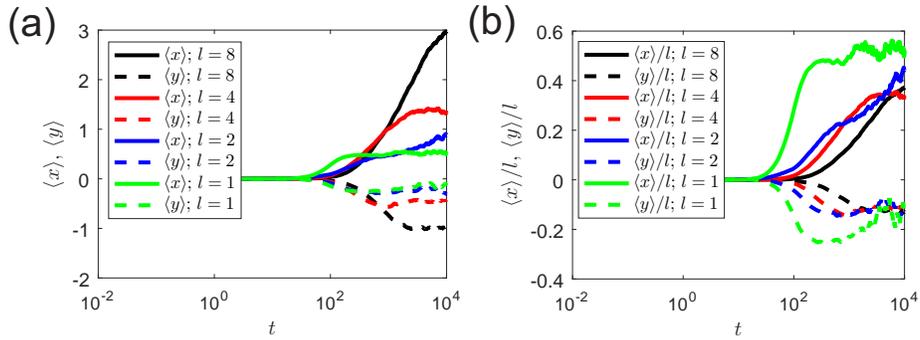}
\caption{\label{fig:saturationComp}(Color Online) (a): MD ($\langle x\rangle$ and $\langle y\rangle$) of CoM versus time based on Fig. \ref{fig:mdCompare}(a)'s results under different triangle edge length $l$. (b): Normalized MD ($\langle x\rangle/l$ and $\langle y\rangle/l$) of CoM versus time based on Fig. \ref{fig:mdCompare}(a)'s results under different triangle edge length $l$.}
\end{figure}
Fig. \ref{fig:saturationComp} (a) shows $\langle x\rangle$ and $\langle y\rangle$ of CoM versus time. Obviously, larger $l$, higher plateau. This is reasonable because the larger triangle, the wider separation between CoM and CoF. Fig. \ref{fig:saturationComp} (b) shows $\langle x\rangle/l$ and $\langle y\rangle/l$ versus time. The normalized MDs converge to the same plateau for different $l$ values, which confirms that the plateau is proportional to and bounded by triangle size.
It also shows the larger the triangle, the longer it takes to achieve the plateau, although given enough time the triangle will eventually equilibrate with the environment. Thenceforth on average the body behaves like a point. Noteworthily, there is a zero size limit. For those triangles that have very small size, the spatial and temporal scale of the convergence becomes negligible. MD of any TP on it would be zero, as predicted by classical theory of Brownian motion.

Because of early stage MD increase and late stage MD plateau, the mean squared displacement (MSD), typically, will first increase fast then converge back to classical Brownian behavior ($\langle\Delta\mathbf{r}^2\rangle\propto t$), exhibiting a crossover behavior~\cite{Ayan}. We would like to refer the readers to works dealing with MSD of different TPs, such as Refs.~\cite{delong,Ayan}. The primary reason why MSD is not chosen here as the signature for discriminating regimes is that the information of direction and orientation is lost in MSD, which is important for our study.

\begin{figure}
\includegraphics[width=0.48\textwidth]{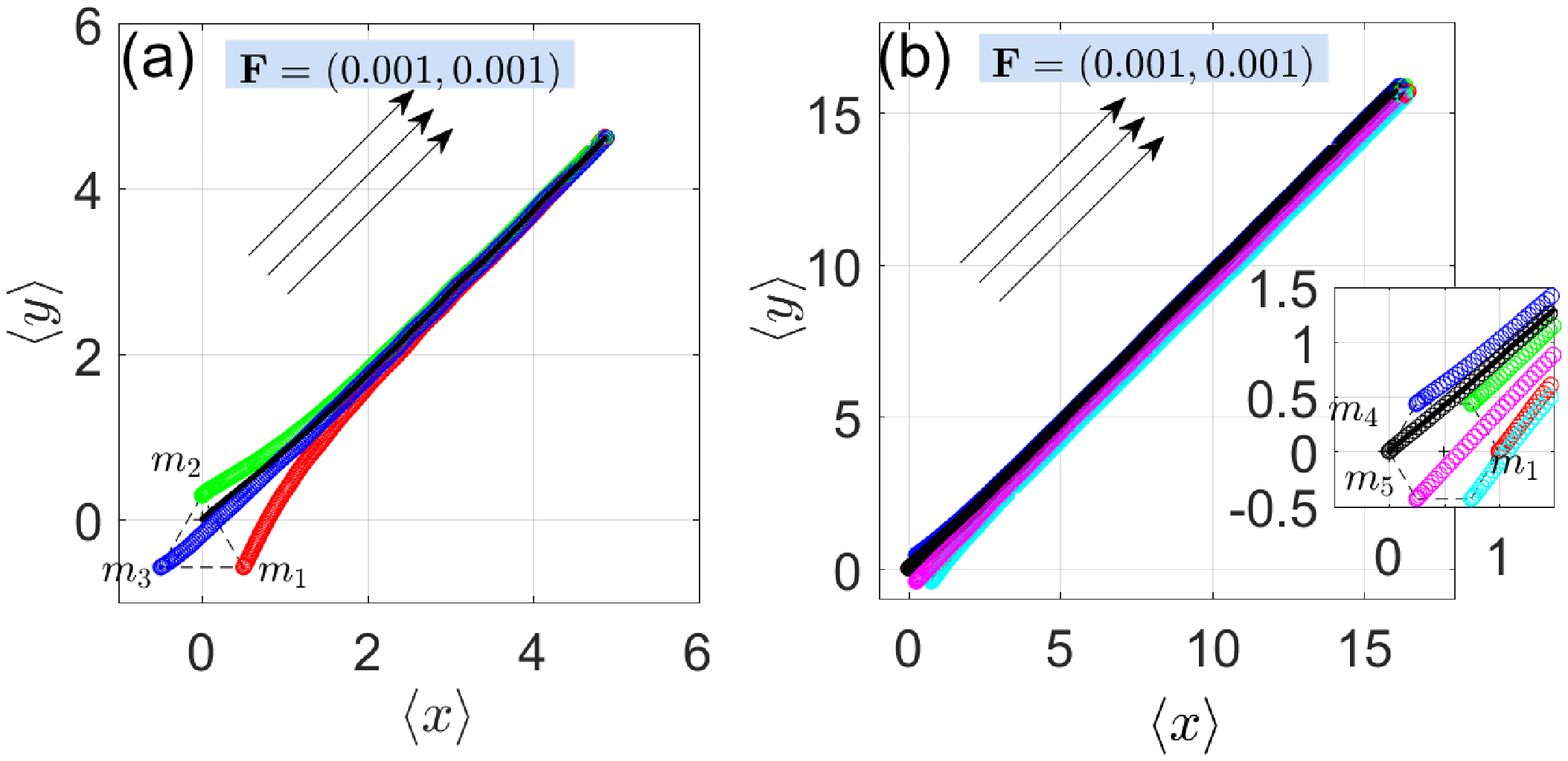}
\includegraphics[width=0.48\textwidth]{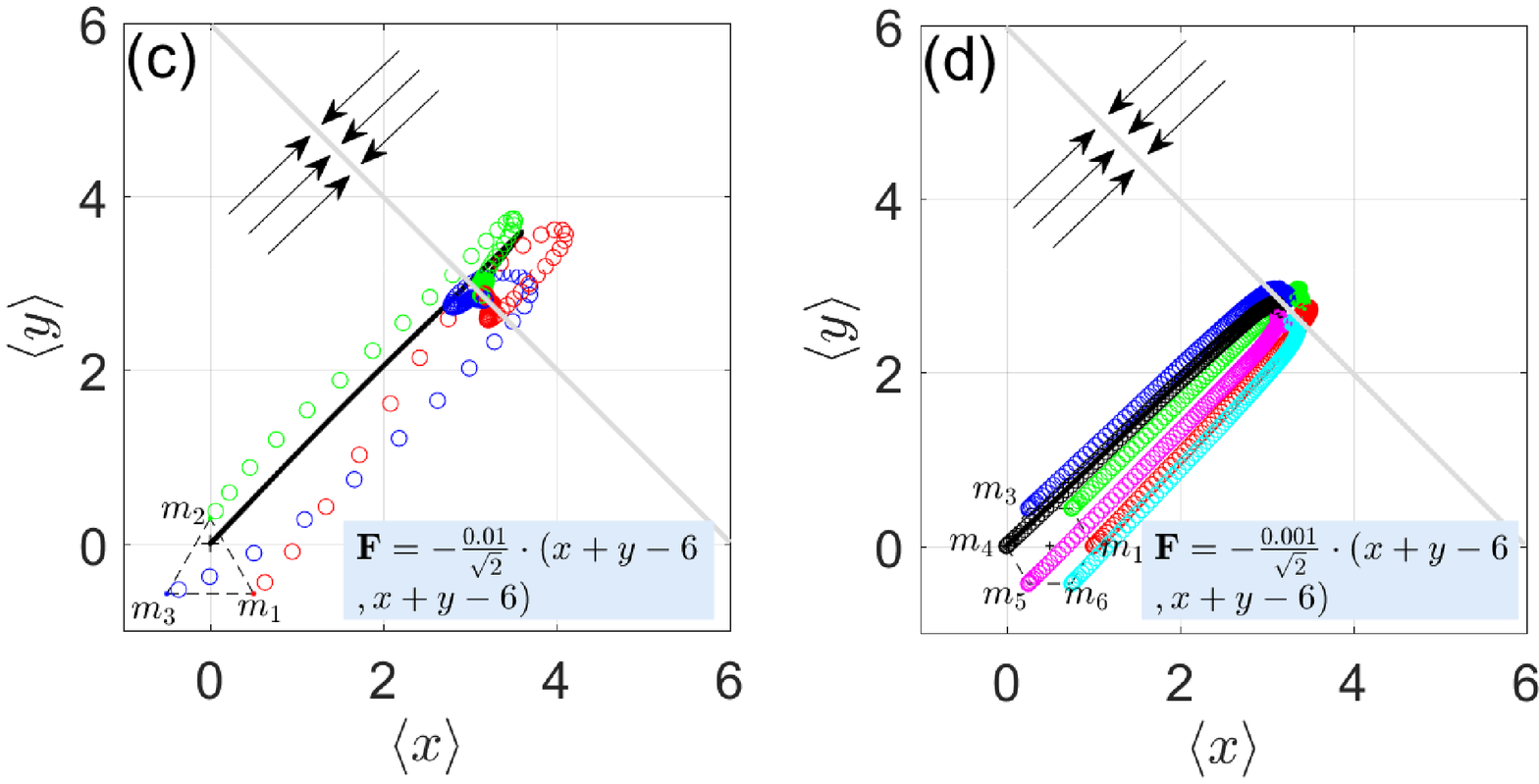}
\caption{\label{fig:convergence_const}(Color Online) Convergence of mean trajectories of vertices on (a) an equilateral triangle (isotropic TRC) and (b) an equilateral hexagon (isotropic non-TRC) under a constant force (c) an equilateral triangle (isotropic TRC) and (d) an equilateral hexagon (isotropic non-TRC) under a harmonic force.}
\end{figure}

Now consider external force $\{\mathbf{F}_i\}_{1\leq i\leq n}$. Two forms of forces are considered here: constant force and harmonic force. In the constant force scenario, each vertex feels the same force no matter where the body is. In the harmonic force case, the forces felt by different vertices are different because they have distinct distances from the valley of harmonic potential. In Fig. \ref{fig:convergence_const} (a), the triangle in Fig. \ref{fig:mdCompare} (b) is subjected to a constant force $\mathbf{F}=(0.001,0.001)$ ($\mathbf{F}_1=\mathbf{F}_2=\mathbf{F}_3=\mathbf{F}$.) The mean trajectories (red for $m_1$, green for $m_2$, blue for $m_3$) of the three vertices and that of the CoM (black solid line) till $t=500$ are shown in the figure. These trajectories converge to a single one, which is the translation of CoF in $\mathbf{F}$'s direction.
As a comparison, in Fig. \ref{fig:convergence_const} (b), the hexagon in Fig. \ref{fig:mdCompare} (d) is forced by a constant force. The figure shows the mean trajectories of six vertices of the hexagon till $t=1300$. The convergence to a single trajectory is also observed.

In Fig. \ref{fig:convergence_const} (c) and (d), the constant force in Fig. \ref{fig:convergence_const} (a) and (b) is replaced by a harmonic force. Depending on the spring strength, mean trajectories can over shoot. However, ultimately they converge to a point on the valley.

The results in this section show that for isotropic body, whether there is external force or not, the mean trajectories converge. They converge to the CoF or the extrapolation of CoF in external force's direction. For anisotropic body, the convergence holds if the body is free.
\section{\label{Align}Alignment of the Mean Trajectories}
\begin{figure}
\includegraphics[width=0.65\textwidth]{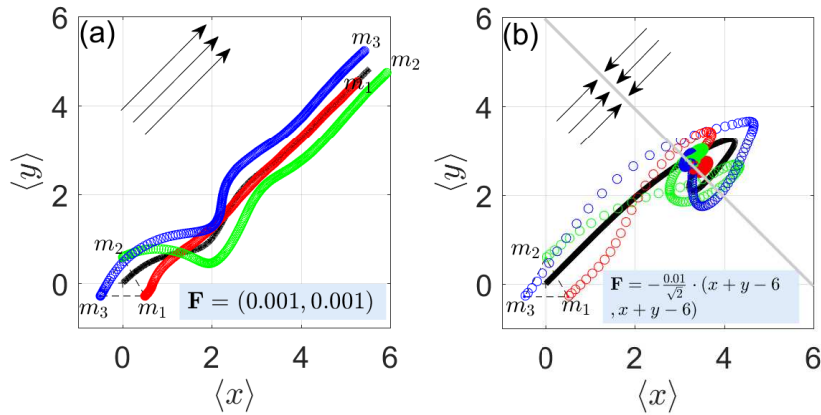}
\caption{\label{fig:alignmentcomp}(Color Online) (a) Alignment of mean trajectories of vertices on an equilateral triangle (anisotropic TRC) under a constant force. (b) The final convergence of mean trajectories of vertices on an equilateral triangle (anisotropic TRC) under a harmonic force.}
\end{figure}

In the alignment regime, the mean trajectories juxtapose each other in parallel. If an anisotropic polygon is subject to external force, the system could fall into the alignment regime. For example, in Fig. \ref{fig:alignmentcomp} (a), the triangle in Fig. \ref{fig:mdCompare} (a) is subjected to a constant force $\mathbf{F}=(0.001,0.001)$ ($\mathbf{F}_1=\mathbf{F}_2=\mathbf{F}_3=\mathbf{F}$). In this case $\xi_1$ is significantly larger than $\xi_2$ and $\xi_3$, and vertex 1 will experience the highest resistance force in the triangle's motion. This is because friction force $\mathbf{f}_i=-\xi_i\mathbf{v}_i=-\xi_i(\mathbf{v}_m+\boldsymbol{\omega}\times\mathbf{R}^m_i)$. Here $\xi_1$ is 27 times larger than $\xi_2$ and $\xi_3$, while the magnitude of $\mathbf{v}_i$ only differs by a factor of $|\boldsymbol{\omega}|\cdot|\mathbf{R}^m_i|$. $|\mathbf{R}^m_i|$ is bounded by triangle size, which is $\sim1$. And according to Fig. \ref{fig:twistregime}, magnitude of $|\boldsymbol{\omega}|$ is smaller than 1 even for very strong force. Therefore the effect of nonuniform $\{\xi_i\}_{1\leq i\leq n}$ is overwhelming and $|\mathbf{f}_1|\gg|\mathbf{f}_2|\simeq|\mathbf{f}_3|$. Consequently, vertices 2 and 3 will be pushed to the front with vertex 1 lagging behind. The polygon reorients to accommodate to the force applied in the $(1,1)$ direction.
After the accommodation, the trajectories continue and they keep the distance from each other.

The situation gets more interesting if we replace the constant force with a harmonic force, as shown in Fig. \ref{fig:alignmentcomp} (b). Similar to (a), the triangle tends to align with the force but before it manages to do so the body shifts to the other side of the potential well. Then it must orient again to the force pointing to opposite direction. Depending on the magnitude of the spring constant, the triangle can touch the center line several times, or just once. Ultimately the triangle resides on the center line where force is zero. Under zero force, the mean trajectories converge. This is consistent with the results in Section \ref{Con}. Consequently, the alignment regime does not appear in this force case. Furthermore, if the spring constant is too small to trap the triangle, the triangle behaves like a free body and undergoes convergence regime as well.
In general, the alignment regime applies when the external force does not frequently change its direction and the force persists long enough such that the polygon has enough time to reorient itself towards the force.

\section{\label{twist}Twist of the Mean Trajectories}
In Fig. \ref{fig:alignmentcomp} (a), before the triangle manages to align with the force there is a period when the force is correcting the triangle's orientation. It is found if the magnitude of the external force rises this regime becomes more salient and unlike the cleanly aligned trajectories they could be twisted and intertwined to form a plait structure. We identify it as the twist regime.
\begin{figure*}
\includegraphics[width=0.98\textwidth]{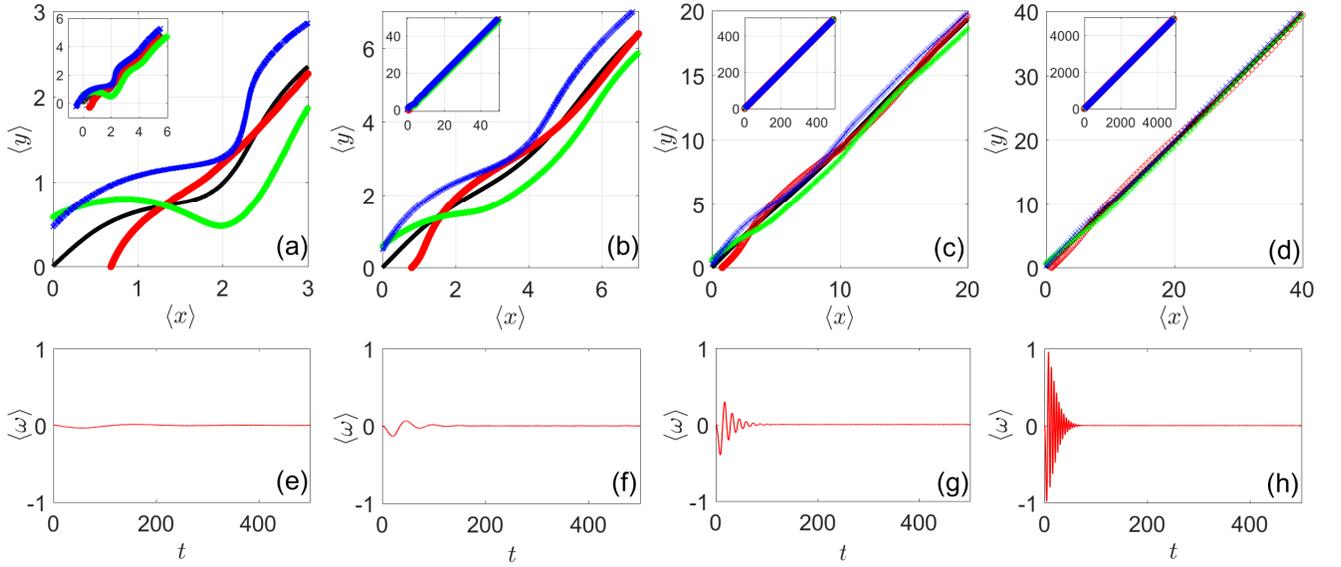}
\caption{\label{fig:twistregime}(Color Online) The twist regime developed before alignment is achieved under different magnitude of the constant external force. (a), (b), (c), (d) show the mean trajectories traced by vertices 1, 2, 3 and CoM (red circle, green asterisk, blue cross and black solid line) based on the simulation in Fig. \ref{fig:alignmentcomp} (a) under forces of $(0.001,0.001)$, $(0.01,0.01)$, $(0.1,0.1)$, $(1,1)$, respectively. The small panel at the top left corner shows the whole trajectory till $t=500$. The major panel shows the snapshot of the twist part of the whole trajectory. As the increase of force magnitude, mean trajectories traced by different vertices become increasingly intertwined. (e), (f), (g), (h) display the ensemble average of the triangle's angular velocity corresponding to (a), (b), (c), (d).}
\end{figure*}

Fig. \ref{fig:alignmentcomp} (a) is selected as the base case. As shown in Figs. \ref{fig:twistregime} (a)-(d), the magnitude of the constant force rises from 0.001, 0.01, 0.1 to 1. As the force strengthens, the twist of the mean trajectories becomes increasingly significant. Figs. \ref{fig:twistregime} (e)-(f) show the ensemble average of angular velocity corresponding to (a)-(d). $\langle\omega\rangle$ reflects the rotation caused by deterministic force. In (a) and (e), the force is small and one witnesses very mild undulation of mean angular velocity which ultimately attenuates to zero, meaning the triangle has reached the stable alignment. The triangle first rotates clockwise (negative $\langle\omega\rangle$), then counterclockwise (positive $\langle\omega\rangle$). When the force rises to 0.01, in (b) and (f) the frequency of switching between clockwise and counterclockwise ramps up. Amplitude of the mean angular velocity also gets higher. Proceeding to 0.1 ((c) and (g)) and 1 ((d) and (h)), both the frequency and amplitude build up. This is because a large force overcorrects the orientation and this overcorrection gets corrected again and again until the triangle reaches the stable orientation. Under the switching between clockwise and counterclockwise rotation, the triangle wiggles around the force's direction and the mean trajectories get twisted and intertwined. Ultimately the triangle aligns with the force.

Note that twist may not necessarily end up in alignment. For example, in the harmonic force case in Fig. \ref{fig:alignmentcomp} (b), the triangle ends up in convergence regime. Therefore what regimes to sample also depends on the form of external force. Under realistic conditions, the potential field could be irregular such that the force frequently changes its direction and magnitude, which may keep the system in the twist regime indefinitely.

\begin{figure}
\includegraphics[width=0.35\textwidth]{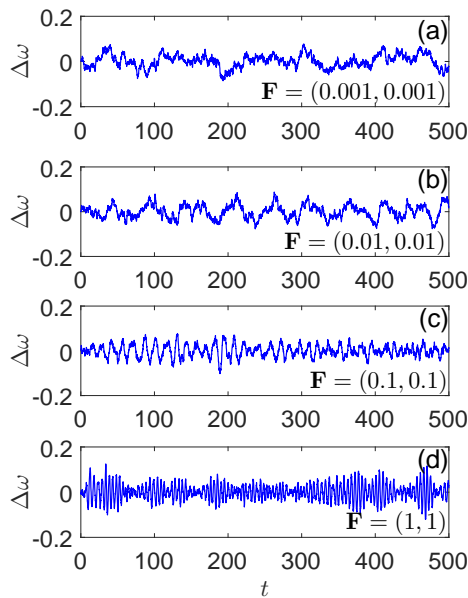}
\caption{\label{fig:frocemagnitude} (Color Online) Instantaneous angular velocity contributed from Brownian source under different magnitudes of external force for anisotropic triangle.}
\end{figure}

One interesting aspect that can be investigated by this section's apparatus is how the Brownian behavior changes as one increases the magnitude of external force. One obtains the Brownian angular velocity by subtracting $\langle\omega\rangle$ from the instantaneous angular velocity $\omega$,
\begin{equation}
\label{eq:bangular}
\Delta\omega=\omega-\langle\omega\rangle.
\end{equation}
$\Delta\omega$ represents contribution from the Brownian source. Typical results for $\Delta\omega$ corresponding to Fig. \ref{fig:twistregime} (e) - (h) are shown in Fig. \ref{fig:frocemagnitude} (a) - (d).
The figure shows that as the magnitude of external force rises, the Brownian fluctuation is excited. After alignment has been achieved (mostly before $t=200$ as shown in Fig. \ref{fig:twistregime}), the Brownian force is still making small rotations of the triangle when $\langle\omega\rangle=0$. Again, the excitation is caused by the correction mechanism. Every now and then the aligned triangle proposes a random rotation, the force rejects and corrects it. And if the force is large, the correction is quick. It looks as if the external force is fueling the Brownian rotation. Numerically it stems from the translation-rotation coupling in the original equations.
Conversely, if we go from large force to small force to zero force, the Brownian fluctuation relaxes to low frequency. Therefore, typically the convergence regime features relaxed Brownian fluctuation.

\section{\label{transition}Transition among convergence, alignment and Twist}
\begin{figure}
\includegraphics[width=0.57\textwidth]{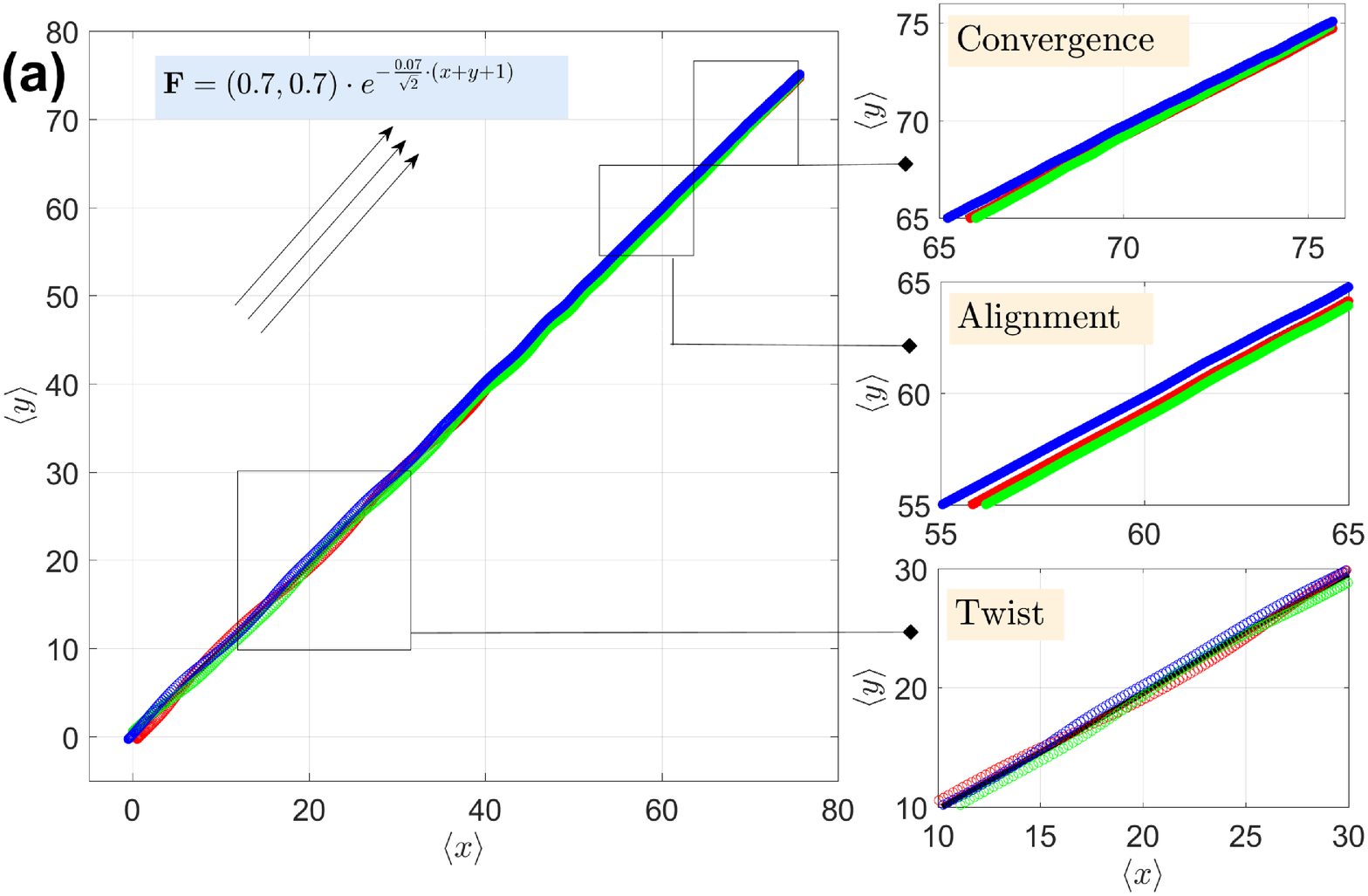}
\includegraphics[width=0.42\textwidth]{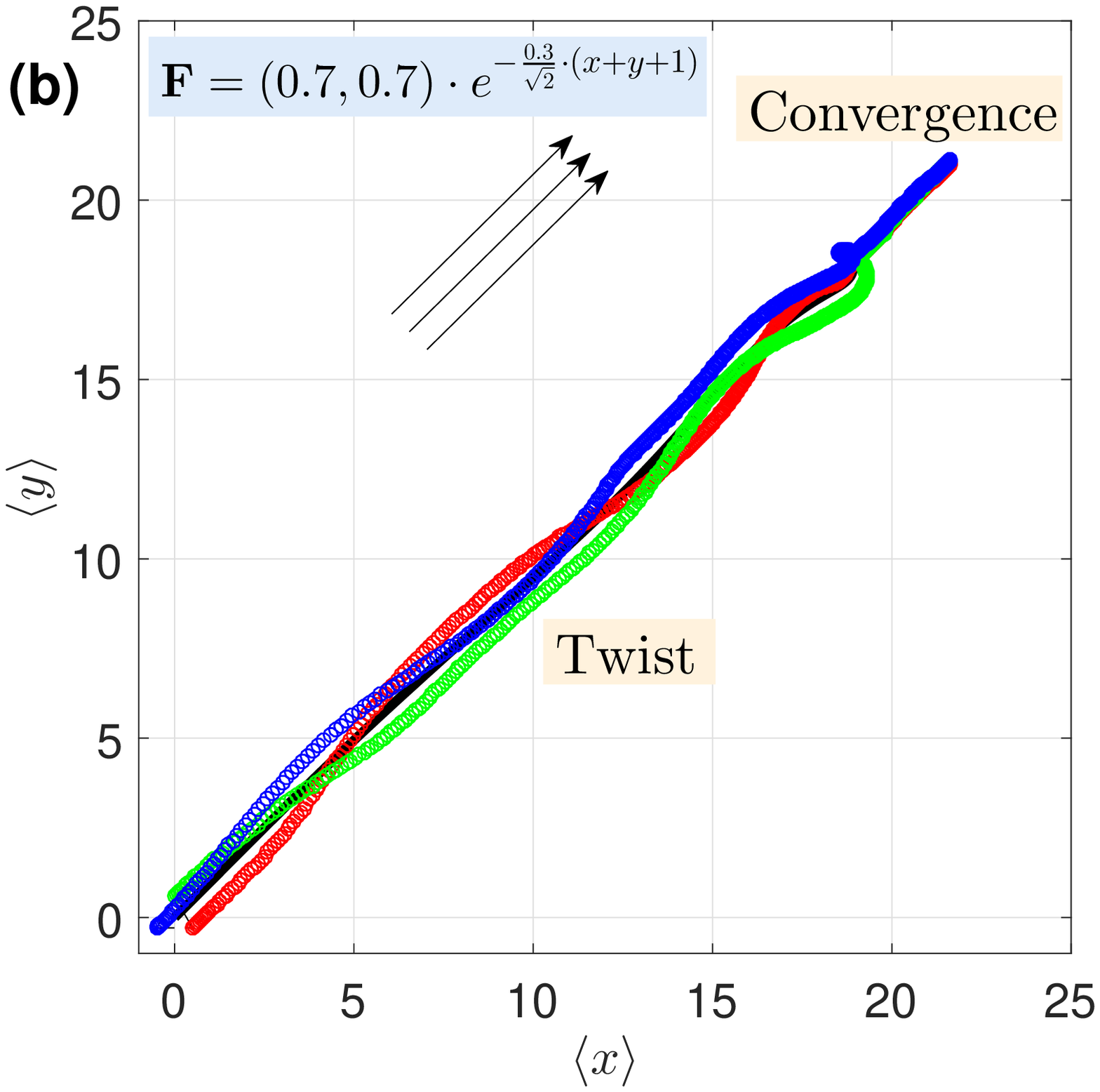}
\caption{\label{fig:tactransition} (Color Online) Mean trajectories traced by vertices of the triangle in Fig. \ref{fig:alignmentcomp} (a) under an exponentially decaying force in space till $t=500$. (a) Small decay rate. (b) High decay rate.}
\end{figure}

As partly discussed in previous sections, polygon can sample and transit from one regime to another. In this section we consider a force that exponentially decays in space, where all the three regimes can be experienced in one path. Based on Fig. \ref{fig:alignmentcomp} (a), the constant force is replaced by an exponentially decaying force with respect to distance in (1,1) direction. The mean trajectories under a small decay rate is shown in Fig. \ref{fig:tactransition} (a).
The triangle experiences sequentially the twist, alignment and convergence regimes until the force decays to zero. If, instead, the decay rate is high, one may only have the twist to convergence transition, as shown in Fig. \ref{fig:tactransition} (b).

\begin{figure}
\includegraphics[width=0.5\textwidth]{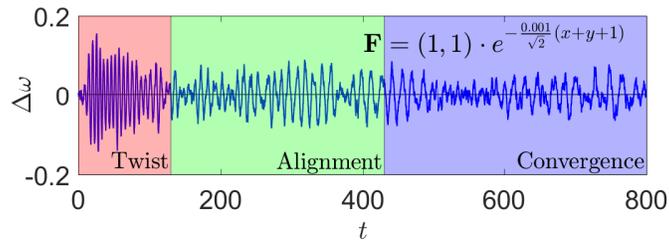}
\caption{\label{fig:tactranomega} (Color Online) The frequency attenuation of instantaneous angular velocity contributed from Brownian source as the system undergoes a twist to alignment to convergence transition under a slowly decaying exponential force.}
\end{figure}

In general, the overall magnitude of force determines the sampling between convergence and nonconvergent regimes (alignment and twist), while the details of the force will determine whether alignment or twist to choose. For example, for free anisotropic body (zero force), convergence rules. Once a nontrivial force is established, all three regimes could be possible depending on the details and idiosyncracies of the field. For instance, as shown above, an anisotropic body ends up in alignment under a constant force whereas it ends up in convergence in harmonic and exponentially decaying potentials.
Since the Brownian rotation is correlated with magnitude of external force (Fig. \ref{fig:frocemagnitude}), if the regime change involves change in force magnitude, one shall expect the modification of Brownian behavior. We arbitrarily select a single triangle from the ensemble and obtain its Brownian angular velocity $\Delta\omega$ till $t=800$ under the context of Fig. \ref{fig:tactransition} with a force that is slowly decaying ($\mathbf{F}=(1,1)\cdot e^{-\frac{0.001}{\sqrt{2}}(x+y+1)}$). The results are shown in Fig. \ref{fig:tactranomega}.
It could be seen that the frequency undergoes attenuation as the regime changes from twist to alignment to convergence.
Sometimes the transition between twist and alignment does not involve change in force's magnitude (such as the constant force case), then there will not be frequency change in such transition.
\section{\label{2to3}From Polygon to Polyhedron}
It is well-known that 3D Brownian motion is different from its 2D version~\cite{Han0,Han1,MUKHIJA}. As discussed in Refs.~\cite{Han0,Han1,MUKHIJA}, 2D and quasi-2D confinement significantly increase friction anisotropy of the particle and impact the anisotropy in diffusion compared to 3D. In other words, from 2D to 3D, one may expect the influence of friction anisotropy to decrease. Therefore it is interesting and meaningful to investigate how the patterns for 2D polygon we found above change when one instead has a polyhedron.

\begin{figure*}
\includegraphics[width=0.98\textwidth]{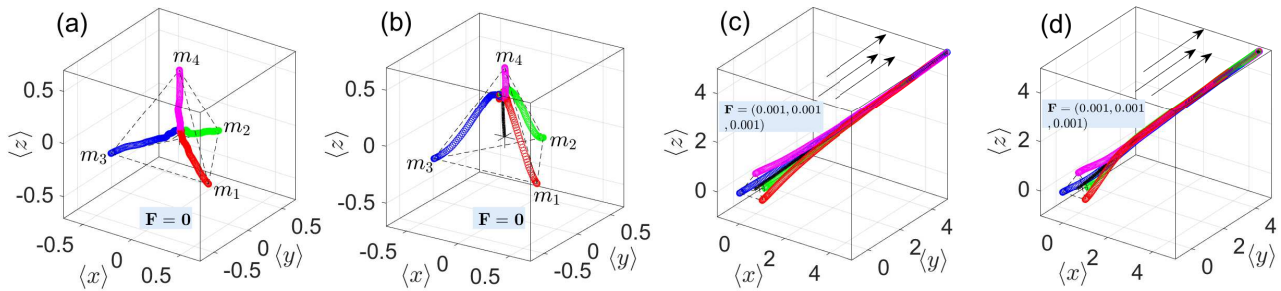}
\caption{\label{fig:convergence_3d}(Color Online) The convergence of mean trajectories of vertices on a tetrahedron (a) of isotropic non-TRC in the absence of eternal force, (b) of anisotropic TRC in the absence of external force, (c) of isotropic non-TRC in the presence of a constant force, (d) of anisotropic TRC in the presence of a constant force. Red, green, blue, magenta circles for trajectories of vertices 1, 2, 3, 4. Black solid line for trajectory of CoM.}
\end{figure*}

As displayed in Fig. \ref{fig:convergence_3d} (a),
we start by considering a simple isotropic equilateral tetrahedron without external force. The initial coordinates for the four vertices 1, 2, 3, 4 are $(\frac{1}{2},-\frac{\sqrt{3}}{6},-\frac{\sqrt{6}}{12})$, $(0,\frac{\sqrt{3}}{3},-\frac{\sqrt{6}}{12})$, $(-\frac{1}{2},-\frac{\sqrt{3}}{6},-\frac{\sqrt{6}}{12})$, $(0,0,\frac{\sqrt{6}}{4})$, respectively. $m_1=m_2=m_3=m_4=1$. In (a) the friction is distributed evenly among vertices as $\xi_1=\xi_2=\xi_3=\xi_4=0.075$. Then by definition the CoM and CoF coincide at $(0,0,0)$. The figure shows the mean trajectories of the four vertices till $t=160$. They converge to the CoF in the end.

In Fig. \ref{fig:convergence_3d} (b), the friction is redistributed as $\xi_1=\xi_2=\xi_3=0.04$ and $\xi_4=0.18$. Now we have an anisotropic tetrahedron with vertex 4 experiencing much more resistance from the medium. Consequently, the new location for CoF is about $(0,0,0.2858)$. The figure shows the mean trajectories of the four vertices and the CoM (black solid line) till $t=160$. In the end they all converge to the CoF. Results in Fig. \ref{fig:convergence_3d} (a) and (b) lead to the same conclusion as 2D system that in the absence of external force the mean trajectories converge regardless of isotropy and anisotropy.

We carry on to subject the isotropic tetrahedron in (a) to a constant external force $\mathbf{F}=(0.001,0.001,0.001)$, shown in Fig. \ref{fig:convergence_3d} (c). This figure shows the mean trajectories till $t=400$. In the end they converge to one single trajectory. This again draws the same conclusion as the 2D system that the mean trajectories converge for isotropic body under external force. As a comparison, Fig. \ref{fig:convergence_3d} (d) shows the results of subjecting the anisotropic tetrahedron in (b) to the constant force. They end up in convergence as well. This differs from 2D system where an anisotropic body under external force can experience the alignment and twist regime. In 3D, the alignment and twist regimes disappear. It could be understood as that the body tries to align itself with the force, but the rotation with respect to the ``alignment axis'' would again lead to convergence. The wiggle by the ``alignment axis'' is also evened out by the extra rotations hence there is no twist regime. Adding more rotational degrees of freedom has made the particle behave more isotropically~\cite{MUKHIJA}.

We summarize the results thus far in Tab. \ref{tab:regimeclass}.
\begin{table}
\caption{Regime classifications for interactions between mean trajectories of multiple tracking points on a Brownian rigid body. (A = Alignment, C = Convergence, T = Twist)}
\label{tab:regimeclass}
\begin{center}
\begin{tabular}{ccccc}
\hline
\hline
 & \multicolumn{2}{c}{2D} & \multicolumn{2}{c}{3D}\\
 & Free & Forced & Free & Forced\\
\hline
Isotropic & C & C & C & C\\
Anisotropic & C & C/A/T & C & C\\
\hline
\hline
\end{tabular}
\end{center}
\end{table}
As the table shows, the convergence regime has an overwhelming dominance in most of the situations. Only for forced anisotropic body in 2D, one might have the alignment and twist regimes which are made possible by translation-rotation coupling. Translation-rotation coupling is also responsible for the modification of Brownian behavior in regime transition. However, such coupling yields to the overwhelming effect of multiple rotational degrees of freedom in 3D space. This is consistent with the results for an ellipsoid where the effect of translation-rotation coupling is much stronger under 2D and quasi-2D confinement~\cite{Han0} compared to 3D.

\section{\label{Conclusion}Conclusions}
We have identified three regimes of interaction, namely, convergence, alignment and twist, between mean trajectories of different tracking points on a Brownian rigid body based on a polygon/polyhedron representation. Depending on the properties of the rigid body and external force, in 2D the body can sample from and transit between these three regimes, while in 3D there is only convergence to sample. And when a body in 2D is transiting between regimes, its Brownian behavior could be modified. The translation-rotation coupling plays a fundamental role in making the nonconvergent regimes (alignment and twist) possible. Otherwise the convergence regime dominates.

Our results show that in most situations, different tracking points on a Brownian rigid body are statistically the same, because of the convergence of mean trajectories. Only for systems in alignment and twist regimes, different tracking points are statistically different, such that an inappropriately chosen tracking point could lead to error in displacement. However, such error is shown to be bounded by the size of the particle. Therefore, if the particle size is relatively small compared to the spatial scale one is interested in, the issue of choosing a tracking point will not be a concern. Nevertheless, in scenarios such as rigid particle sedimentation near a surface, the spatial scale is reduced to be commensurate with particle size, then there could be a special choice of tracking point that best estimates long-time transport coefficient~\cite{delong}.

The result that only convergence regime survives from 2D to 3D suggests Brownian behaviors of rigid body in bulk medium and under confinement are quite different. Confinement in space limits the number of rotational degrees of freedom and allows more anisotropic behaviors, while increasing the number of rotational degrees of freedom makes the particle behave much more isotropically~\cite{Han0,Han1,MUKHIJA}. It is somewhat surprising that our extremely simplified model has captured such dramatic change during dimensional change. Indeed, there remain more details to uncover in future works about the interactions between anisotropy, translation-rotation coupling and spatial confinement in Brownian dynamics.
\section*{APPENDIX: The Semi-Analytical Solution of Mean Displacement of CoM}
This appendix presents the semi-analytical solution to mean displacement (MD) of CoM in the absence of external forces.

When $\{\mathbf{F}_i\}_{1\leq i\leq n}=\{\mathbf{0}\}$, from Eq. (\ref{eq:gen-eqn}), one obtains
\begin{equation}
\frac{d\mathbf{v}_m}{dt}=-\frac{\sum_{i=1}^{n}\xi_i}{\sum_{i=1}^{n}m_i}\mathbf{v}_m-\boldsymbol{\omega}\times\frac{1}{\sum_{i=1}^{n}m_i}\sum_{i=1}^{n}\xi_i\mathbf{R}^m_i+\frac{1}{\sum_{i=1}^{n}m_i}\sum_{i=1}^{n}\delta\mathbf{F}_i.
\tag{A.1}
\end{equation}
Integrating Eq. (A.1) and using $\Delta\mathbf{r}_m=\int_0^t\mathbf{v}_mdt'$, we arrive at ($\mathbf{v}_m(0)=\mathbf{0}$ applied) an equation for displacement $\Delta\mathbf{r}_m$,
\begin{equation}
\frac{d}{dt}\Delta\mathbf{r}_m=-\frac{\sum_{i=1}^{n}\xi_i}{\sum_{i=1}^{n}m_i}\Delta\mathbf{r}_m-\frac{1}{\sum_{i=1}^{n}m_i}\int_0^t\boldsymbol{\omega}\times(\sum_{i=1}^{n}\xi_i\mathbf{R}^m_i)dt'
+\frac{1}{\sum_{i=1}^{n}m_i}\sum_{i=1}^{n}\int_0^t\delta\mathbf{F}_idt'.
\tag{A.2}
\end{equation}
Taking ensemble average of Eq. (A.2), one gets the Langevin equation for MD $\langle\Delta\mathbf{r}_m\rangle$,
\begin{equation}
\frac{d}{dt}\langle\Delta\mathbf{r}_m\rangle=-\frac{\sum_{i=1}^{n}\xi_i}{\sum_{i=1}^{n}m_i}\langle\Delta\mathbf{r}_m\rangle-\frac{1}{\sum_{i=1}^{n}m_i}\Big\langle\int_0^t\boldsymbol{\omega}\times(\sum_{i=1}^{n}\xi_i\mathbf{R}^m_i)dt'\Big\rangle.
\tag{A.3}
\end{equation}
By Eq. (\ref{eq:prop2}), the integral in Eq. (A.3) could be carried out
\begin{equation}
\frac{d}{dt}\langle\Delta\mathbf{r}_m\rangle=-(\frac{\sum_{i=1}^{n}\xi_i}{\sum_{i=1}^{n}m_i})\langle\Delta\mathbf{r}_m\rangle-\frac{1}{\sum_{i=1}^{n}m_i}\sum_{i=1}^{n}\xi_i\langle\mathbf{R}^m_i\rangle
+\frac{1}{\sum_{i=1}^{n}m_i}\sum_{i=1}^{n}\xi_i\mathbf{R}^m_i(0).
\tag{A.4}
\end{equation}
Applying the initial condition $\Delta\mathbf{r}_m(0)=\mathbf{0}$, solution to Eq. (A.4) is
\begin{equation}
\langle\Delta\mathbf{r}_m\rangle=-\frac{1}{\sum_{i=1}^{n}m_i}\int_0^t\Big(\sum_{i=1}^{n}\xi_i(\langle\mathbf{R}^m_i(t')\rangle-\mathbf{R}^m_i(0))\Big) e^{-\frac{\sum_{i=1}^{n}\xi_i}{\sum_{i=1}^{n}m_i}(t-t')}dt'.
\tag{A.5}
\end{equation}
The second integral can be taken out and Eq. (A.5) simplifies to
\begin{equation}
\langle\Delta\mathbf{r}_m\rangle=\frac{\sum_{i=1}^{n}\xi_i\mathbf{R}^m_i(0)}{\sum_{i=1}^n\xi_i}(1-e^{-\frac{\sum_{i=1}^{n}\xi_i}{\sum_{i=1}^{n}m_i}t})
-\frac{1}{\sum_{i=1}^{n}m_i}\int_0^t\Big(\sum_{i=1}^{n}\xi_i\langle\mathbf{R}^m_i(t')\rangle\Big) e^{-\frac{\sum_{i=1}^{n}\xi_i}{\sum_{i=1}^{n}m_i}(t-t')}dt'.
\tag{A.6}
\end{equation}
Eq. (A.6) can be expressed in a more concise way. Let's introduce the vector $\mathbf{S}_F^m$, i.e., the vector joining from the CoM to the CoF.
The $\sum_{i=1}^{n}\xi_i\langle\mathbf{R}^m_i(t')\rangle$ term can be rewritten as
\begin{equation}
\sum_{i=1}^{n}\xi_i\langle\mathbf{R}^m_i(t')\rangle=\langle\sum_{i=1}^{n}\xi_i(\mathbf{S}_F^m(t')+\mathbf{R}^f_i(t'))\rangle=(\sum_{i=1}^{n}\xi_i)\langle\mathbf{S}_F^m(t')\rangle.
\notag
\end{equation}
This is true because of Eq. (\ref{eq:prop1}). Therefore Eq. (A.6) reduces to
\begin{equation}
\langle\Delta\mathbf{r}_m\rangle=\mathbf{S}_F^m(0)(1-e^{-\frac{\sum_{i=1}^{n}\xi_i}{\sum_{i=1}^{n}m_i}t})-\frac{\sum_{i=1}^{n}\xi_i}{\sum_{i=1}^{n}m_i}\int_0^t\langle\mathbf{S}_F^m(t')\rangle e^{-\frac{\sum_{i=1}^{n}\xi_i}{\sum_{i=1}^{n}m_i}(t-t')}dt'.
\tag{A.7}
\end{equation}

\begin{figure}
\includegraphics[width=0.4\textwidth]{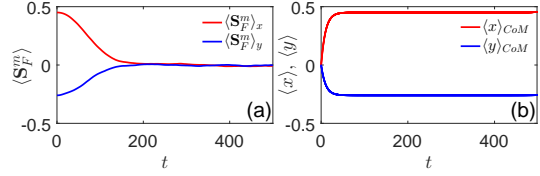}
\caption{\label{fig:aa}(Color Online) (a) The behavior of $\langle\mathbf{S}_F^m\rangle$ versus time. (b) The behavior of $\langle\Delta\mathbf{r}_m\rangle$ versus time.}
\end{figure}
This result shows that $\langle\Delta\mathbf{r}_m\rangle$ is biased towards CoF until it saturates. This is only a semi-analytical result because we do not have an analytical solution for $\langle\mathbf{S}_F^m\rangle$, which is determined by the angular velocity $\omega$. The translation-rotation coupling in Eq. (\ref{eq:gen-eqn}) makes it difficult to obtain an analytical solution for $\omega$. However, based on the case in Fig. \ref{fig:mdCompare} (a), we numerically compute the behavior of $\langle\mathbf{S}_F^m\rangle$ versus time. The result is shown in Fig. \ref{fig:aa} (a).
With the numerical solution for $\langle\mathbf{S}_F^m\rangle$, we can substitute it into Eq. (A.7) and obtain the MD of CoM. The result is shown in Fig. \ref{fig:aa} (b). This result indicates that the contribution from the integral part in Eq. (A7) is negligible and the MD is almost determined by the first part of Eq. (A7) - a saturated exponential growth, which agrees with the experimental and theoretical results of a boomerang colloidal particle study~\cite{Ayan}.

\bibliography{apssamp}


\end{document}